\documentclass[aps, prd, reprint, groupaddress]{revtex4-1}
\usepackage{graphicx}
\usepackage{dcolumn}
\usepackage{bm}
\usepackage{hyperref}
\usepackage{latexsym}
\usepackage{amsmath}

\usepackage{amssymb, amsthm}
\usepackage{mathrsfs}

\begin{document}

\title{Search for a stochastic gravitational-wave background
using a pair of torsion-bar antennas}%

\author{Ayaka Shoda$^1$}
\email{shoda@granite.phys.s.u-tokyo.ac.jp}
\author{Masaki Ando$^1$}
\author{Koji Ishidoshiro$^2$}
\author{Kenshi Okada$^1$}
\author{Wataru Kokuyama$^1$}
\author{Yoichi Aso$^1$}
\author{Kimio Tsubono$^1$}%
\affiliation{$^1$ Department of Physics, The University of Tokyo, Hongo 7-3-1, Tokyo 113-0033, Japan}
\affiliation{$^2$ Research Center for Neutrino Science, Tohoku University,
Sendai 980-8578, Japan} %

\begin{abstract}
We have set a new upper limit on the stochastic gravitational-wave background
using two prototype torsion-bar antennas (TOBAs).
A TOBA is a low-frequency gravitational-wave detector with bar-shaped test
masses rotated by the tidal force of gravitational waves. 
As a result of simultaneous 7-hour observations with TOBAs in
Tokyo and Kyoto in Japan, our upper limit with a confidence level of 95\% is
$\Omega_{\rm gw}h_0^2 < 1.9 \times 10^{17}$ at 0.035 -- 0.830 Hz, where $h_{0}$
is the Hubble constant in units of 100 km/s/Mpc and $\Omega_{\rm gw}$ is the
gravitational-wave energy density per logarithmic frequency interval in units of
the closure density.
We successfully updated the upper limit and extended the
explored frequency band.
\end{abstract}

\maketitle

\section{Introduction}
Detecting a stochastic gravitational-wave background (SGWB) is one of the most
ambitious targets in gravitational-wave (GW) astronomy.
A SGWB is a superposition of GWs produced in
the early Universe and GWs emitted from astronomical
sources with amplitudes too small to be resolved.
Direct observation of a SGWB is fundamentally important to
understanding how the universe evolved because a GW can carry information about 
the universe before the epoch of the last scattering of the cosmic microwave
background (CMB) photons due to its high transparency.
Revealing the frequency dependance as well as the amplitude of a SGWB
will strongly constrain many inflation models since the phase transition of the vacuum or
preheating of the universe will produce peaks in the power spectrum density
(PSD) of a SGWB \cite{SGWBmodels, Nakayama2008}.

Several upper limits on a SGWB have been established by observations.
For example, interferometric GW detectors, LIGO and Virgo, have set an upper
limit of around 200 Hz \cite{LIGO_UL}.
LIGO and a resonant bar detector ALLEGRO were used together to establish an
upper limit of around 900 Hz.
Two cryogenic resonant bars Explorer and Nautilus have searched for a SGWB
at 900 Hz \cite{cryo_bar}.
Other than those, a pair of synchronous interferometers have placed an upper
limit at 100 MHz \cite{akutsu}.
At lower frequencies, upper limits have been set at $10^{-6} - 10^{-3}$ Hz by
Doppler tracking of the Cassini spacecraft \cite{Cassini}, and at $10^{-9} -
10^{-7}$ Hz by pulsar timing that measured the fluctuations in pulse arrival
times from PSR B1855+09 \cite{Maggiore_PR}.
COBE set an upper limit at $10^{-18} - 10^{-16}$ Hz by observing the CMB
\cite{stochastic_allen}.
Regarding indirect evidence, the helium-4 abundance resulting
from big-bang nucleosynthesis (BBN) \cite{Maggiore_PR} and measurement of the
CMB and matter power spectra \cite{CMB} set a constraint on the integrated cumulative energy density
of a SGWB. 

In addition to these observations, a torsion-bar antenna (TOBA) has opened the
frequency band that other GW detectors cannot access \cite{TOBA}.
A TOBA is a GW detector with bar-shaped test masses, which is sensitive at low
frequencies such as 0.1 -- 1 Hz even on the ground.
 We have already set the first upper limit at 0.2 Hz of $\Omega_{\rm gw} h_0^2
 \leq 8.7 \times 10 ^{17}$ using a 20 cm scaled prototype TOBA
 \cite{singleTOBA_UL}.
However, this upper limit was set using a single detector, and it is difficult
to distinguish a SGWB signal from noise using only one detector because the
waveform of a SGWB is random and unpredictable.
Therefore, simultaneous observations with
multiple detectors are required for a direct search for a SGWB.
In addition, the signal-to-noise ratio is improved by the square root of
the observation time because the uncorrelated noises at two separated places
would be suppressed.
Therefore, we searched for a SGWB by the cross-correlation analysis with
two prototype TOBAs.
As a result, we were able to update the upper limit on a SGWB and extend the
explored frequency region.
We present the search procedure and results in this paper.

\section{Observation}
\begin{figure}[bt]
\includegraphics[width=14pc]{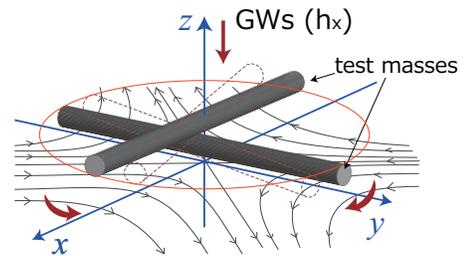}
\caption{\label{fig:concept} (color online) A conceptual drawing of a TOBA.
Two test mass bars are positioned orthogonally and rotate differentially
due to the tidal force of the GWs.}
\end{figure}
A TOBA \cite{TOBA} is a GW detector composed of two bar-shaped orthogonal test
masses that rotate differentially due to the tidal force caused by GWs as shown in Fig.
\ref{fig:concept}. Its angular fluctuation $\theta(t)$ obeys the equation of motion:
\begin{equation}
I \ddot{\theta}(t) + \gamma \dot{\theta}(t) + \kappa \theta(t) = \frac{1}{4}
\ddot{h}_{ij} q^{ij},
\end{equation}
where $I, \gamma, \kappa, h_{ij}$, and $q^{ij}$ are the moment of inertia, the
dissipation term, the spring constant in the rotational degree of freedom, the
amplitude of a GW, and the quadropole moment of the test mass, respectively.
This results in a simple equation $\tilde{\theta}(f) = q^{ij}h_{ij}/2I$ above
the resonant frequency $f_0 = \sqrt{\kappa/I}/2\pi$.
TOBA is fundamentally sensitive at lower frequencies, even on the ground.
The low resonant frequency in the rotational degree of freedom $f_0$ on the
order of a few millihertz makes the test mass free at 0.1 -- 1 Hz.
The small rotational seismic motion also contributes to the good sensitivity at
low frequencies.

We have developed small prototype TOBAs in Tokyo and Kyoto.
The longitude and latitude of the two sites are $139.76^{\circ},
35.71^{\circ}$ (Tokyo) and $135.78^{\circ}, 35.03^{\circ}$ (Kyoto).
Each prototype TOBA has one 20 cm test mass bar that is magnetically levitated
by a pinning effect between a magnet attached at its top and a superconductor at
the top of the vacuum tank \cite{ishidoshiro2010}.
The angular fluctuation of the test mass is read by a Michelson
interferometer; the two laser beams split by a beam splitter hit mirrors
attached to both ends of the test mass.
Therefore, the difference in the two beam path lengths is proportional to the
rotation angle of the test mass.
Both test masses are oriented in a north-to-south direction and controlled by
coil-magnet actuators so that the fringes of the interferometers are kept in the middle.

\begin{figure}[bt]
\includegraphics[width=18pc]{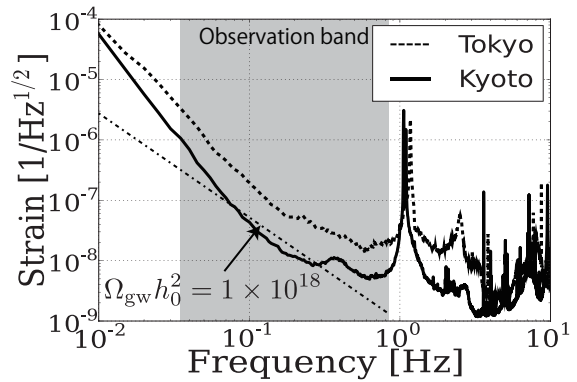}
\caption{\label{fig:strain} Equivalent strain noise spectra of two
detectors. Dashed and solid lines indicate spectra of Tokyo and Kyoto
respectively.
The sensitivity in the Tokyo data is lower than that in the Kyoto data due to
there being many glitches in the data from Tokyo. These glitches were removed
during data selection.}
\end{figure}

We performed simultaneous observations with the prototypes for about 7 hours
from 21:21 JST to 4:48 JST on October 29, 2011.
Figure \ref{fig:strain} shows the equivalent strain noise spectra of the two
detectors averaged over the whole observation time.
The strains are limited by seismic noise coupled from translational
motion at frequencies higher than 0.5 Hz and by the magnetic coupling noise at
lower frequencies \cite{singleTOBA_UL}.
The noise level is relatively high because there are several glitches which
are clearly not a GW signal.

\section{Analysis}
The amplitude of a SGWB is usually characterized by the dimensionless quantity
$\Omega_{\rm gw}$ defined as
 \begin{equation}
\Omega_{\rm gw}=\frac{f}{\rho_{\rm c}}\frac{d\rho_{\rm gw}}{df},
\end{equation}
where $\rho_{\rm c}=3c^2 H_0^2 /8\pi G$ is the critical energy density of the
universe and $\rho_{\rm gw}=\langle \dot{h}_{ij}\dot{h}^{ij} \rangle c^2/32\pi
G$ is the energy density of a SGWB.
Then, the PSD of a SGWB can be written as
\begin{equation}\label{eq:stogw}
S_{\rm gw}(f) = \frac{3H_0^2}{10\pi^2}f^{-3}\Omega_{\rm gw}(f),
\end{equation}
where $H_0$ is the Hubble constant.
Here we assume that a SGWB is stationary, isotropic, unpolarized and Gaussian,
and $\Omega_{\rm gw} = {\rm constant}$ in our observation band.

In order to search for a SGWB, we take the general cross correlation between the
two detector outputs in strain $s_1(t)$ and $s_2(t)$:
\begin{eqnarray}\label{eq:Y}
Y & = & \int_{-T/2}^{T/2} dt \int_{-T/2}^{T/2} dt' s_1(t)s_2(t')Q(t-t') \nonumber \\
 & \sim & \int_{-\infty}^{+\infty} df \tilde{s}_1^*(f) \tilde{Q}(f) \tilde{s}_2(f).
\end{eqnarray}
Here, a tilde notates Fourier transformed functions. $T$ and $\tilde{Q}(f)$ are
the observation time and a filter function chosen so that the
signal-to-noise ratio of $Y$ is maximized.
The expected value of $Y$ depends only on a SGWB since the noise in each
detector is uncorrelated.
Considering Eq. (\ref{eq:stogw}), $\tilde{Q}(f)$ is written as
\begin{equation}
\tilde{Q}(f) = C \frac{\gamma(f)}{P_1(f) P_2(f) f^3},
\end{equation}
where $P_i(f)$ is the power spectrum density of the $i$th detector's output, and $C$
is a normalization factor set in order to give $\langle Y \rangle = \Omega_{\rm
gw}h_0^2 T$.
$h_0$ is the normalized Hubble constant defined as $h_0 = H_0/100{\rm km}/{\rm
sec}/{\rm Mpc}$. 
$\gamma (f)$ is the overlap reduction function that represents the difference
between the two detector responses.
In this case, $\gamma(f)$ is almost unity below 10 Hz since distance between the
two sites is about 300 km which is short compared to the GW wavelength and the
bars were oriented in the same direction.
Please refer to \cite{Maggiore_book} for further details.

\begin{figure}[bt]
\includegraphics[width=20pc]{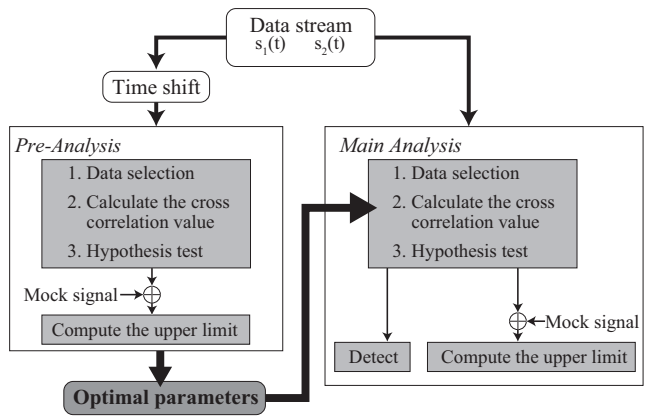}
\caption{\label{fig:flowchart}Flowchart of the analysis.
The preanalysis and main analysis are similar except that the
preanalysis is performed using the time-shifted data to tune the
parameters.}
\end{figure}

A flowchart of the data analysis is shown in Fig. \ref{fig:flowchart}.
The data stream is divided into $N$ segments with 50\% overlap, and fast Fourier
transformation was performed using a Hanning window.

After the data are transformed into $\tilde{s}_{iJ}(f)$, the data for
the $J$th segments of $i$th detector's output, we removed segments in which
the noise was obviously not a GW signal and was large enough to affect the
result.
The removed segments were selected according to the band-limited root mean
square (rms).
We did not use the main observation band for data selection to avoid
unintentionally removing a signal.
Instead, we calculated the rms below 0.05 Hz and above 1 Hz as an indicator of
the data selection in order to remove segments where the magnetic coupling noise
or seismic noise was large.

Next, the cross-correlation value $Y/T_{\rm seg}$, which corresponds to
$\Omega_{\rm gw} h_0^2$, is calculated at each surviving segment.
Here, $T_{\rm seg}$ is the length of the segments.
We limited the integration frequency band to where $\tilde{Q}(f)$ was biggest,
since $\tilde{Q}(f)$ is largest at frequencies where the sensitivity to a SGWB is the best.

We then judged that a SGWB signal is present if $\langle Y \rangle /T_{\rm
seg}$, where $\langle Y \rangle$ is the average cross-correlation value for the
segments, is larger than the detection threshold.
This test is based on the Neyman-Pearson criterion \cite{NeymanPearson}.
A detection threshold depends on the probability distribution of $\langle Y
\rangle/T_{\rm seg}$ without a signal and the false alarm
rate $\alpha$.
The probability distribution, which reflects the background, is
estimated from observed data.
A histogram of $\langle Y_{JK} \rangle /T_{\rm seg}$, which is the
cross correlation between $\tilde{s}_{1J}(f)$ and $\tilde{s}_{2K}(f) (J \neq K,
K \pm 1)$, is proportional to the probability distribution without a signal
because there is no correlation between two segments whose time differs by more
than the time constant of a target signal, even though a signal is present in
the original data stream.
Note that we did not put the correlation between the segments next to each
other so the correlation in overlapped time does not affect the result.
Also, the data are considered to be sufficiently stationary through the
observation such that the correlation values of the segments with large
separation of time will not cause the histogram to differ from the true
probability distribution.
Thus, the detection threshold $z_{\alpha}$ is set so that the integral
of the probability distribution from $z_{\alpha}$ to $\infty$ is equal to $\alpha$.
When $\langle Y \rangle /T_{\rm seg}$ calculated with time adjusted data is
larger than $z_{\alpha}$, a signal is present.
Here, the correlation value is not necessarily positive because we did not align
the plus and minus of the signal in this analysis.

If a SGWB signal is not detected, an upper limit is set by mock signal
injection based on a Bayesian method.
We injected a mock signal into the real data and searched for the mock signal as
mentioned above.
The mock signal is the random data stream created by filtering a Gaussian number
sequence to have the frequency dependence shown in Eq.
(\ref{eq:stogw}).
We repeated the mock signal search many times and set the $\beta$
confidence level upper limit as the amplitude of the mock signal detected with a
probability of $\beta$.

Here, there are arbitrary parameters such as the length of the segments, the
ratio of the removed segments, and the bandwidth of the integrated frequency
region.
It is necessary to search for the optimal parameters using the
actual data since these parameters depend on the data quality.
We chose the parameters that maximize a ''pseudo'' upper limit derived by
the same analysis as the main analysis with time-shifted data (preanalysis).
In this paper, we shifted 2,000 seconds to tune the parameters.

\section{Result}
\begin{figure}[bt]
\includegraphics[width=18pc]{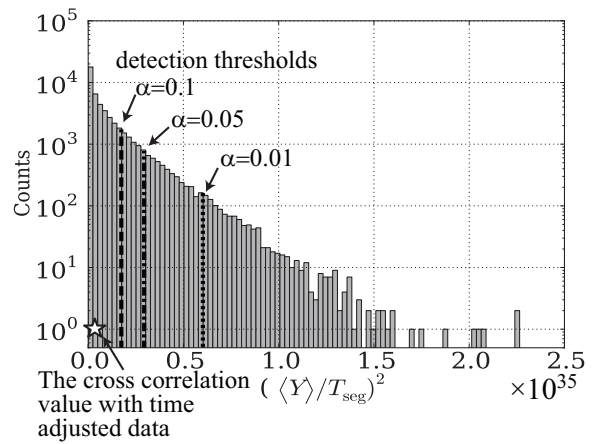}
\caption{\label{fig:histogram}Histogram of $\left( \langle Y \rangle
/T_{\rm seg} \right)^2$. 
Dashed and broken lines are detection thresholds with false alarm
rates of 10\%, 5\%, and 1\%. The star shows the cross-correlation value
with time-adjusted data.}
\end{figure}
As a result of parameter tuning, data were divided into segments of 200
seconds, and 10\% of the segments were removed by the data selection.
In total, 206 segments were used to calculate $\langle Y \rangle$.
We set the analyzed frequency bandwidth as 0.8 Hz and integrated from 0.035
to 0.830 Hz to calculate $Y$.

Using these parameters, the detection threshold with a false alarm rate of 5\%
is $|z_{0.05}| = 1.7 \times 10^{17}$ according to the histogram of $\left( \langle
Y \rangle / T_{\rm seg} \right)^2$ shown in Fig.\ref{fig:histogram}.
The cross-correlation value calculated with the time adjusted data is $\langle
Y \rangle /T_{\rm seg} = -5.9 \times 10^{16}$.
Therefore, we concluded that no SGWB signal was detected in our data.
As a result of the mock signal injection, our 95 \% confidence level upper limit
without systematic errors is $\Omega_{\rm gw} h_0^2 < 1.7 \times 10^{17}$.

The systematic error arises mainly from the overlap reduction function and
the calibration.
The error in the overlap reduction function occurs because the direction of the
test mass is not strictly aligned.
We estimate the error in the relative angle of the test masses to be $\pm
5^{\circ}$.
Then, the error in the overlap reduction function is 10\%.
The main reason for the calibration error is the uncertainty of the
beam spots on the mirrors.
The angular fluctuation of the test mass $\theta$ is derived as $\delta l
/L$, where $\delta l$ and $L$ are the change in the beam path length of the
interferometer and the distance between the centers of the two mirrors
attached at both ends of the test mass.
The calibration error appears in $\theta$ because the beam spots are not always
on the centers of the two mirrors, which means that $L$ has an error.
These errors are estimated to be 10\%.
Therefore, the total conservative error is $10 + 10 = 20$\% and our upper limit
with 95\% confidence level including the error is $\Omega_{\rm gw}h_0^2 \leq
1.9 \times 10^{17}$.


{\it Discussion and Future Plans} ---
Considering the integrated upper limits, $\Omega_{\rm gw} h_0^2$ has already
been constrained by the BBN or CMB measurements at 0.035 -- 0.830 Hz.
However, these upper limits are only on cosmological SGWBs, and not on
astronomical SGWBs.
Our result is the first to set the upper limit using a direct search
constraining both SGWBs in this frequency band.

The result derived from cross-correlation analysis is expected to be better
by a factor of $\gamma_{\rm rms} \sqrt{T_{\rm obs} \Delta_{\rm BW}}$
compared to the result derived using a single detector with the same
sensitivity, where $\gamma_{\rm rms}, T_{\rm obs}$, and $\Delta_{\rm BW}$ are the
observation time, the bandwidth of the integration of the cross correlation,
and the rms of $\gamma(f)$ over that bandwidth, respectively
\cite{Abbott2004}.
Though $\gamma_{\rm rms}\sqrt{T_{\rm obs} \Delta_{\rm BW}} \sim 1 \times 10^2$
with our configuration of $\gamma_{\rm rms} \sim 1, \Delta_{\rm BW} = 0.8,
T_{\rm obs} \sim 2 \times 10^4 {\rm s}$, our result is only about 4 times better than the previous result.
This is because we could not achieve the sensitivity obtained in 2009 \cite{singleTOBA_UL}.
The poor alignment would induce large coupling noises.
Still, we successfully updated the upper limit and extended the explored
frequency region due to the uncorrelated noise reduced by the cross-correlation
analysis.

For SGWB detection, there is a need to upgrade the setup.
Cross correlation analysis using one-year observation data
with two TOBAs with a 10 m scaled configuration \cite{TOBA} should detect a
SGWB with $\Omega_{\rm gw}h_0^2 > 10^{-8}$.
However, it is difficult to achieve such sensitivity at the next upgrade
due to several technical problems, such as magnetic coupling noise, seismic coupling noise, thermal noise, and the Newtonian noise.
Therefore, we are now constructing a second prototype, Phase-II TOBA.
Two orthogonal test masses and an optical bench will be suspended by wires in order to reduce the magnetic coupling noise and common mode noise.
In addition to introducing a vibration isolation system, we will monitor the
test masses in all the degrees of freedom and diagonalize the signal so that the
motion in the pendulum mode will not couple to the signal.
The suspension wires will be cooled down for thermal noise reduction.
The Newtonian noise \cite{Saulson1984, Hughes1998} would affect the sensitivity
of TOBA in the same way as for the interferometric GW detectors and is estimated
to be observed below 0.1 Hz with Phase-II TOBA. We will try to test to subtract
it using sensor arrays \cite{Driggers2012} because the Newtonian noise cannot be shielded out.
Its fundamental sensitivity will be about $h \sim 1 \times 10^{-15} /\sqrt{\rm Hz}$ at
1 Hz in strain.
Using one-year simultaneous observation data with this sensitivity, the upper
limit on a SGWB will be improved to $\Omega_{\rm gw}h_0^2 \leq 1$.
Moreover, we will introduce a new method for deriving multiple independent data
observed with different directivity.
Combining the cross-correlation analysis and this technique, TOBA will have the
advantage of mapping a full-sky map of an astronomical SGWB, as well as
searching for a cosmological SGWB.

{\it Conclusion} ---
We performed simultaneous 7-hour observations with two
prototype TOBAs in Tokyo and Kyoto and searched for a SGWB using
cross-correlation analysis.
A SGWB signal was not detected and the new 95\% confidence upper
limit is $\Omega_{\rm gw} h_0^2 \leq 1.9 \times 10^{17}$ at 0.035 -- 0.830 Hz.
This is the first experimental demonstration of a direct SGWB search using
cross-correlation analysis with two TOBAs.
The results allowed an update of the upper limit and extended exploration of the frequency band.

This work was supported by JSPS KAKENHI Grants No. 24244031, No. 24$\cdot$7531,
and No. 25610046.

\bibliographystyle{apsrev4-1}
\bibliography{TOBA_refs}

\end{document}